\documentclass[notitlepage,superscriptaddress,longbibliography,10pt]{revtex4-1}
\setcitestyle{super}
\usepackage[colorlinks=true, allcolors=blue]{hyperref}

\usepackage{etoolbox}		
	\patchcmd{\subsection}{\centering}{\hspace*{2em}\raggedright}{}{}
	\patchcmd{\subsubsection}{\centering}{\hspace*{0.25em}\raggedright}{}{}
	\patchcmd{\subsubsection}{\normalfont\small\itshape}{\normalfont\small\itshape\bfseries}{}{}
\makeatletter				
	\renewcommand*{\thesection}{\arabic{section}}
	
	\renewcommand*{\p@subsection}{}
	
	\renewcommand*{\p@subsubsection}{}
\makeatother

\usepackage{amsmath,amssymb,esint}				
\usepackage{lmodern}							
\usepackage{graphicx}
\graphicspath{{./Figures/}}

\usepackage{xcolor}

\usepackage{lineno}

\newcommand{\be}{\begin{equation}}
\newcommand{\ee}{\end{equation}}
\newcommand{\BE}{\begin{eqnarray}}
\newcommand{\EE}{\end{eqnarray}}
\newcommand{\nn}{\nonumber}
\newcommand{\bra}{\left\langle}
\newcommand{\ket}{\right\rangle}
\newcommand{\avg}[1]{\bra{#1}\ket}
\newcommand{\Da}{\mbox{Da}}
\newcommand{\bv}{\mathbf{v}}

\begin{document}

\title{Fast flowing populations are not well mixed}

\author{Francisco Herrer\'{i}as-Azcu\'{e}}
	\email{francisco.herreriasazcue@manchester.ac.uk}
	\affiliation{Theoretical Physics, School of Physics and Astronomy, The University of Manchester, Manchester M13 9PL, United Kingdom}
\author{Vicente P\'{e}rez-Mu\~{n}uzuri}
	\email{vicente.perez@cesga.es}
	\affiliation{Group of Nonlinear Physics, Faculty of Physics, University of Santiago de Compostela E-15782 Santiago de Compostela, Spain}
\author{Tobias Galla}
	\email{tobias.galla@manchester.ac.uk}
	\affiliation{Theoretical Physics, School of Physics and Astronomy, The University of Manchester, Manchester M13 9PL, United Kingdom}


		\begin{abstract}

In evolutionary dynamics, well-mixed populations are almost always associated with all-to-all interactions; mathematical models are based on complete graphs. In most cases, these models do not predict fixation probabilities in groups of individuals mixed by flows. We propose an analytical description in the fast-flow limit. This approach is valid for processes with global and local selection, and accurately predicts the suppression of selection as competition becomes more local. It provides a modelling  tool for biological or social systems with individuals in motion.

		\end{abstract}


\maketitle


Population dynamics describes the changes of the composition of a group of individuals over time. Broadly speaking, there are two modelling approaches. One involves well-mixed populations, implying an all-to-all interaction. This is contrasted with structured populations, or populations on networks. Mathematically, the interaction network of well-mixed populations is often assumed to be a `complete graph' (see e.g. \cite{Ohtsuki2006,Nowak2010,Hindersin2015,Santos2006}), i.e., a network in which interaction links exist between any two individuals at all times. In the context of epidemics, for example, an infection event can affect any of the susceptible individuals in the population; in evolutionary dynamics, it indicates that competition occurs between all members of the population. This effectively means that there is no spatial structure at all, or at least that interaction is sufficiently long-range that spatial structure is not relevant for the evolutionary process.

In this work, we ask if and when populations in a flow are well mixed. Assume we place a population of discrete individuals in a container, bacteria immersed in a fluid perhaps \cite{Lapin2006,Sokolov2007,Venail2008,Leinweber2017a}, and stir the system. One would naturally think that a well-mixed system can be obtained in this way, provided the stirring is sufficiently strong, ergodic and that one waits long enough. 

Our analysis shows, however, that the predictions of the conventional theory for well-mixed populations do not always capture the outcome of evolutionary processes in stirred environments. We use computer simulations to study the outcome of an evolutionary process in a finite population immersed in different flows. Specifically, we focus on the situation in which a single mutant invades an existing population of wildtypes. We then ask when the theory for well-mixed populations quantitatively predicts its chances of success. In the language of population genetics, we study the probability of fixation \cite{Ewens2004}.

We find that the answers to these questions are subtle, and depend very much on the precise concept of what a `well-mixed' population is. Our results show that the validity of the conventional theory is not primarily a question of the nature or speed of the stirring; instead, it is determined by the interaction range and the type of evolutionary process. As a consequence we think the term `well-mixed', which suggests external stirring, needs to be used with care.

We present an analytical approach to stirred populations, in which we abandon the assumption that the interaction graph is complete at all times. Instead, we rely on a broader definition: a population is well mixed if every pair of individuals interacts with the same probability\cite{Lieberman2005,Nowak2006,Hauert2007}. At any one time particles take positions in space and compete within an interaction radius. This situation is not necessarily described by a complete graph. However, if the population is stirred at sufficiently high rates, and if the flow is such that it `mixes' all parts of the system, particles take random positions in space at each evolutionary step. The possible interaction partners of a given individual are then effectively sampled uniformly from the entire population, and any two individuals are equally likely to interact. This process can be described analytically, and fixation probabilities can be obtained. In contrast to the conventional theory for well-mixed populations, this method accurately reproduces simulation results for stirred systems.

We show that the conventional theory is only valid for processes in which selection is global, i.e., it occurs between \emph{all} individuals in the population.  The method presented here, on the other hand, is also valid for local selection. Since the details of the evolutionary dynamics in real-world systems are rarely known with certainty, this flexibility makes our approach relevant for the modelling of experiments where the interacting individuals are in motion.

\section{Well-mixed populations}\label{sec:WM}

It is fair to say that there is a consensus on what constitutes a well-mixed population in mathematical models of evolutionary dynamics. In order to illustrate this, we focus on stochastic dynamics in finite populations, and use a discrete-time Moran process \cite{Moran1958}. We choose a death-birth update; this is sufficient to present our results. Other mechanics of evolution could equally well be considered (see Supplemental Material).

The population consists of $N$ individuals; we assume that its size is constant over time. Each individual can either be a mutant or a wildtype. The state of the population at any point in time is characterised by the number of mutants, $m$; the number of wildtypes is $N-m$. In a non-spatial setting there is nothing else to know about the state of the population. In the Moran model, evolution occurs through combined death-birth events. In each event, one individual is picked at random, regardless of its type, and is removed from the population. One of the remaining individuals is then selected for reproduction and generates an offspring. Selection is based on fitness, and the offspring is of the same species as its parent. We assume frequency-independent selection, and set the fitness of wildtype individuals to one. The invading mutant has fitness  $r$, which can be smaller than one (for disadvantageous mutations) or greater than one (for advantageous mutations). For $r=1$ the process is neutral.

The dynamics proceed through a sequence of death-birth events until the mutants have either gone extinct ($m=0$), or reached fixation ($m=N)$. When either has occurred the dynamics stop, as there is only one type of individual left in the population. We focus on the probability, $\phi$, for a single invading mutant to reach fixation. Using the theory of Markov processes, an explicit mathematical expression can be found for the fixation probability (see e.g. \cite{Ewens2004,Gardiner2003,Nowak2006a,Traulsen2009}). The fixation probability depends on the fitness of the mutant species, $r$, and the population size, $N$. For the death-birth process one has \cite{Kaveh2015,Hindersin2015}
	\be
	\phi = c\,\frac{1-r^{-1}}{1-r^{-cN}},
	\label{eq:CG_fixprob}
	\ee
where $c=(N-1)/N$ for the process described above. We will refer to this as the prediction of the conventional theory for well-mixed populations. If asked how to calculate fixation probabilities in well-mixed populations, most evolutionary theorists would likely point to results such as the one in Eq.~(\ref{eq:CG_fixprob}). The details of the mathematical expression may vary for different processes (for example $c=1$ if self-replacement is included or if the selection is global), but they are all derived from the assumption of an all-to-all interaction at all times.

It is more difficult to determine if a particular biological or social system is well mixed. The term is used in the literature without much specificity. Common characterisations include the requirement that `all pairs of individuals interact with the same probability' (see e.g. \cite{Lieberman2005, Nowak2006, Hauert2007, Durrett1994}), but how precisely this is to be interpreted is often not said. For example, do all individuals have to be able to interact with each other at all times? Or it is sufficient if all pairs interact with equal frequency over time?

Not much is usually said to explain how the complete interaction graph leading to Eq.~(\ref{eq:CG_fixprob}) would arise. The term attributed to this formalism ---evolutionary dynamics in `well-mixed' populations---at the very least suggests that this all-to-all interaction can be achieved through some type of external stirring or agitation. It is this assumption that we challenge in this paper.

\section{Models of stirred populations}\label{sec:Stirring}
In order to model the effects of external stirring we assume that the population is subject to a continuous-time flow, moving the individuals around in space \cite{Biology2003, Pigolotti2012, Groselj2015, Pigolotti2012a, Karolyi2000, Karolyi2005,  Cowen2000, Krieger2017, Young2001}. For example, one may imagine a population of bacteria in an aqueous environment, which is being shaken or stirred mechanically \cite{Lapin2006, Sokolov2007, Venail2008, Leinweber2017a}. We focus on two-dimensional systems; this is sufficient to develop the main points we would like to make. 

Each evolutionary death-birth event of the Moran process is executed as follows. First, one individual in the population is chosen at random for removal; each individual is chosen with equal probability $1/N$. Then, its `neighbours' (individuals within interaction range $R$) compete to reproduce and fill the vacancy.  This competition is decided by fitness: assuming that $n$ wildtypes and $m$ mutants compete, the probability that the reproducing individual is a wildtype is $n/(n+mr)$, and the probability that a mutant reproduces is $mr/(n+mr)$. The offspring is created with the same type as the parent, and is placed at the position of the individual that has been removed.

It remains to say how often evolutionary events take place relative to the timescale of the flow. In-line with the literature we will characterise this by the so-called Damk\"ohler number, $\Da$ (see e.g. \cite{Sandulescu2007,Neufeld2002,Young2001,Galla2016}). In our simulations, one evolutionary event occurs every $1/(N\Da)$ time units. If $\Da$ is very large, the flow is slow compared to evolution. The extreme case $\Da\to\infty$ describes the `no-flow' limit; on the timescale of the evolutionary dynamics, the positions of the individuals are then static. A very small value of $\Da$ ($\Da\ll 1$) indicates that the flow is fast compared to evolution. If the conventional theory for well-mixed populations (Sec. \ref{sec:WM}) is to apply to populations in a flow, then one would expect it to be in this limit.
	
\begin{figure*}[tb!!]
	\center
	\includegraphics[width=0.8\textwidth]{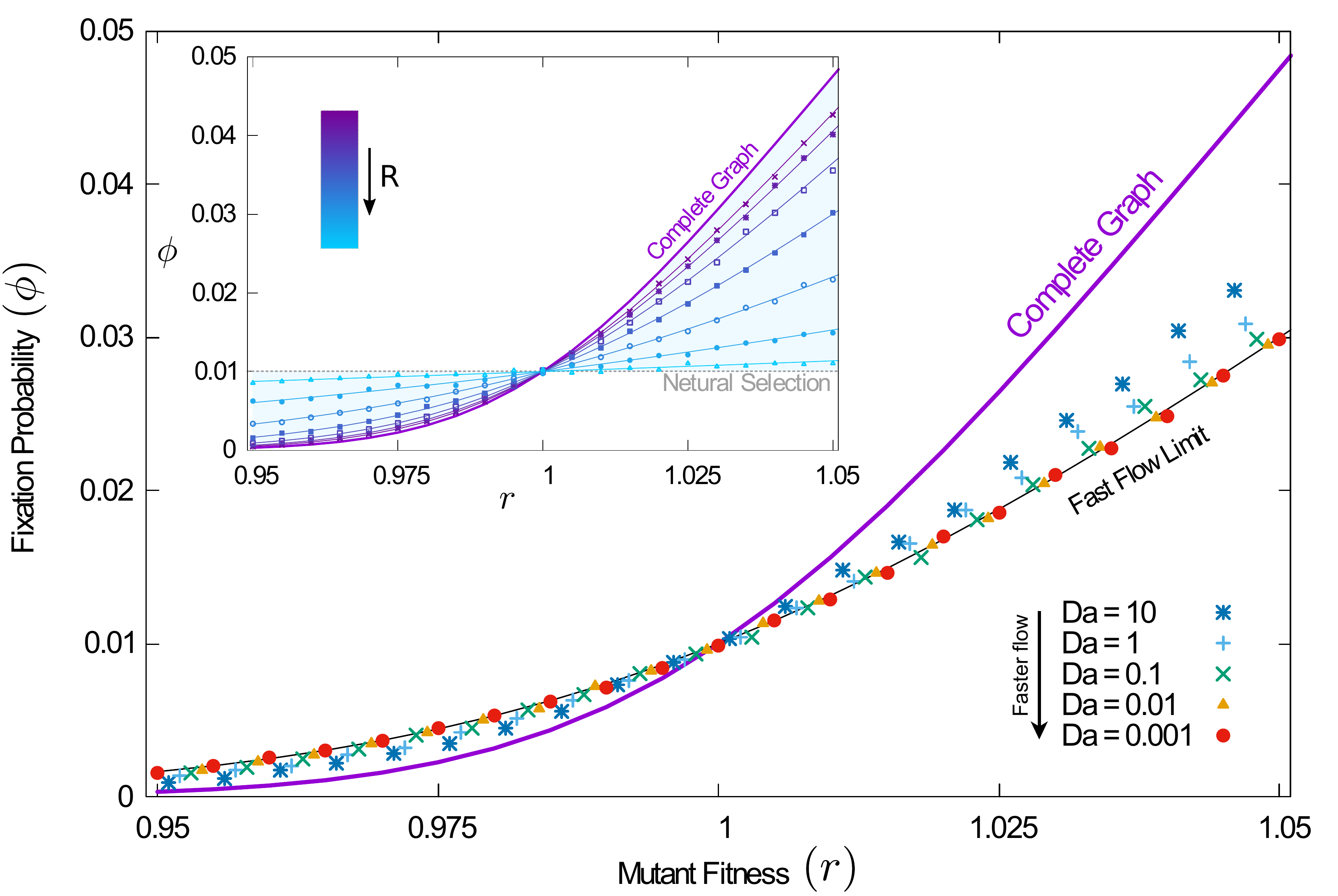}	
	\caption{\textbf{Fixation probability of a single mutant in a population stirred by a parallel-shear flow.} 
				The conventional theory for well-mixed systems [Eq.~\eqref{eq:CG_fixprob}] is shown as a thick purple line. 
				Markers represent simulation results. In the main panel, these are shown for different Damk\"ohler numbers. Reducing $\Da$ increases the flow speed relative to the evolutionary process. The thin continuous lines represent results from the analytical approach for fast flows [Eq.~\eqref{eq:Approx}]. 
				The inset shows simulations for different interaction radii. Smaller interaction ranges makes selection increasingly more local, and the fixation probability approaches that of neutral selection, $1/N$, shown for reference (dashed gray line). (Population size $N=100$; $R=0.1$ in main panel; $\Da=0.1$ in the inset; interaction radius varies from $R=0.025$ to $R=0.175$ in inset. See Supplemental Material for a description of the flow.)
				}
	\label{fig:PShear}
	\end{figure*}

We performed simulations of the evolutionary process in a parallel-shear flow \cite{Ottino1989,Neufeld2010}, which leads to chaotic motion (see Supplemental Material for details on the flow). Results are shown in the main panel of Fig. \ref{fig:PShear}. The thick purple line is the prediction of the conventional theory for well-mixed systems. The markers represent simulation results for different Damk\"ohler numbers. For fixed mutant fitness, the data suggests that the fixation probability  approaches a limiting value as $\Da$ is decreased (the flow made faster). However, this limiting value is not the one predicted by Eq.~(\ref{eq:CG_fixprob}). This indicates that the conventional well-mixed theory is not applicable, even for fast chaotic flows.

\begin{figure*}[t!!]
	\center
	\includegraphics[width=1\textwidth]{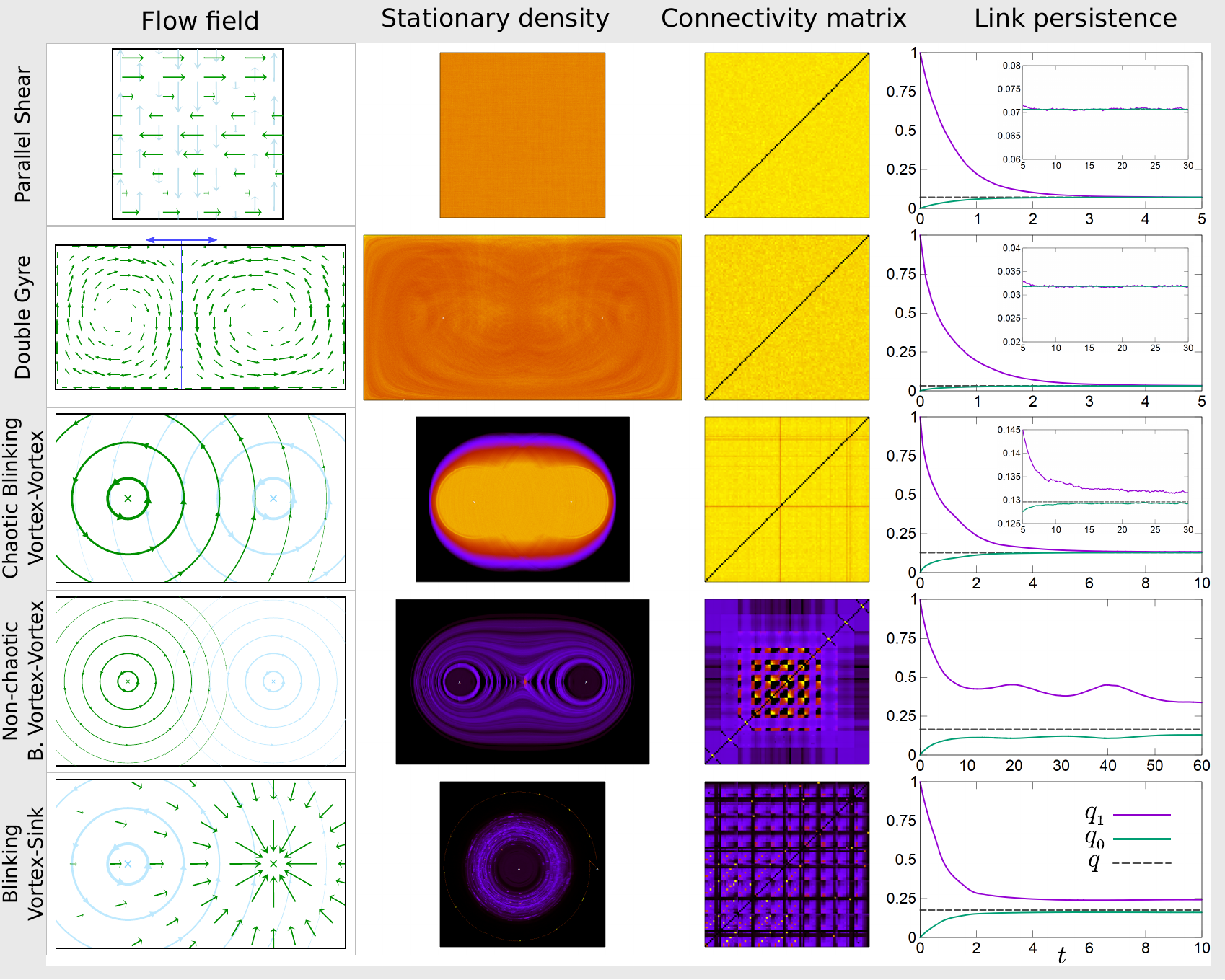}
	\caption{\textbf{Mixing properties of different planar flows.} 
				The first column shows a graphical representation of the flow field for a selection of two-dimensional flows (see Supplemental Material). For periodic flows, green arrows represent the velocity field during one half period, and blue arrows in the other half period. 
				The second column shows the stationary density of particles in space, as measured from simulations. 
				The fraction of time each pair of particles spend within interaction radius from each other is shown as a connectivity matrix in the third column. Results are from simulations. 
				The fourth column shows the measured link persistence, $q_1(t)$, as well as $q_0(t)$ and the asymptotic connectivity $q$ (see text). Convergence of $q_1$ and $q_0$ to a common value $q$ indicates that the flow mixes the system. }
	\label{fig:Table}
	\end{figure*}

\bigskip
We next return to the commonly used description of a well-mixed population, and determine whether any pair of individuals interact with the same probability. Labelling particles and tracking them as the flow proceeds, we determine (for each pair of particles) the proportion of time they are within interaction range from each other. This yields a symmetric connectivity matrix, shown in  Fig. \ref{fig:Table}. Results indicate that, averaged over time, the parallel-shear flow meets the verbal criterion of good mixing; each individual is equally likely to interact with any other.

\begin{figure*}[t!!]
	\center
	\includegraphics[width=0.75\textwidth]{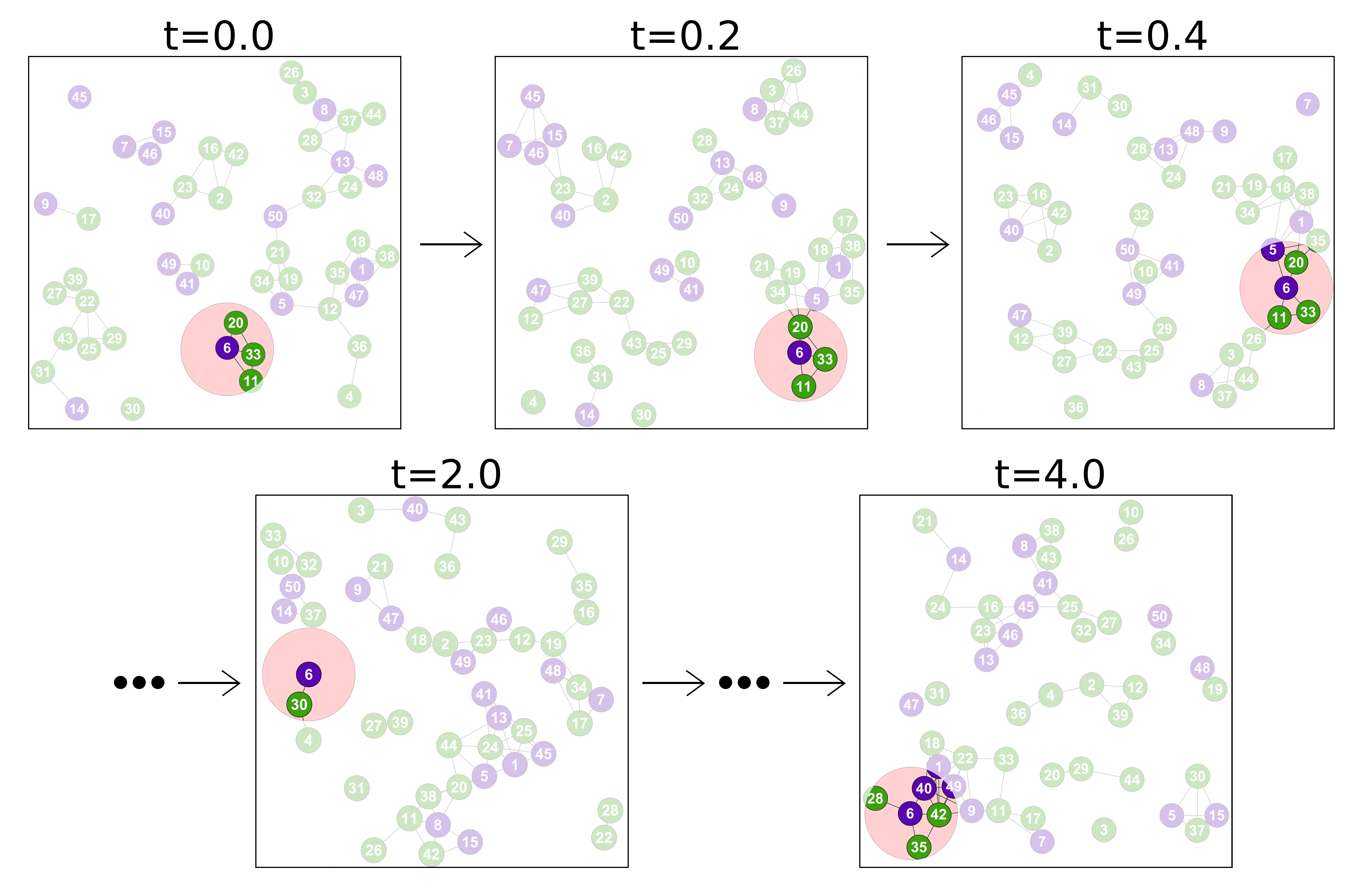}
	 
	\caption{\textbf{Set of neighbours of an individual at different moments in time.} 
				The illustration shows the position of a set of particles as they are moved by the flow. 
				We highlight the time-dependent set of neighbours of one particle. 
				The sets of neighbours remain correlated in the frames shown in the upper row.
				In the lower row, however, the sets of neighbours are uncorrelated from frame to frame. 
				}
	\label{fig:Sequence}
	\end{figure*}

Now, consider the sequence of times at which a specific individual participates in evolutionary events, i.e., it is chosen to compete or to be replaced. In order for a system to be `well-mixed' it is reasonable to require that the set of neighbours at the time of an event is uncorrelated from that at earlier events. If this is not the case, the system has not been `mixed' between the two events. We illustrate this in Fig. \ref{fig:Sequence}. The top row shows snapshots taken at short intervals, and demonstrates that the sets of neighbours of a particle remain correlated from one frame to the next. If evolutionary steps were to happen on these timescales the system cannot be said to be well mixed. If, on the other hand, evolutionary events occur with lower frequency, the neighbours of a particle at the time of an event are uncorrelated to those at earlier events. This is illustrated in the lower row.

To characterise this in more detail, we have measured the probability, $q_1(t)$, that two particles who were within interaction radius at time $t_0$ are also connected at time $t_0+t$. In the stationary state, this is independent of $t_0$. We refer to this quantity as the link persistence. We also measured the probability that two particles who are not connected at an earlier time are connected $t$ units of time later, and denote this quantity by $q_0(t)$. Results are shown in Fig. \ref{fig:Table}.

We write $q$ for probability that two randomly selected individuals are within interaction radius of each other, and refer to this as the connectivity. If a flow mixes the population well we expect that, eventually, the set of neighbours of a particle becomes independent of its earlier neighbours. Then $q_0(t)$ and $q_1(t)$ both tend to $q$. Simulations indicate that this is the case for the parallel-shear flow (see Fig. \ref{fig:Table}). This again confirms that the flow is mixing.

The timescale $t_x$ on which this asymptotic regime is reached can be obtained from the simulation data for the parallel-shear flow. As a broad order-of-magnitude estimate we use $t_x=5$. We expect that mixing occurs on this timescale. The stationary particle density is uniform for this flow and, therefore, the stationary value of $q$ can readily be calculated. We expect $q=\pi R^2/A$, where $A$ is the total area of the two-dimensional system and $R$ the interaction radius. In the example shown in Fig. \ref{fig:Table} we have $A=1$ and $R=0.15$, consistent with the stationary value of $q\approx 0.07$ observed in the figure.

\section{Analytical description}\label{sec:Analytic}

If the typical time elapsing between evolutionary events involving a fixed particle is larger than $t_x$, we can assume that the neighbours of the particle are uncorrelated to those in earlier events. Using this, an analytical description can be constructed. In any evolutionary event, one particle is chosen at random for removal. The neighbours of this particle are obtained by randomly sampling the entire population; each particle is in the neighbourhood of the focal individual with probability $q$. Those neighbours then compete to fill the vacancy. This allows us to derive rates with which mutants replace wildtype individuals or vice-versa. From these rates we then compute the probability for a single mutant to reach fixation. Details can be found in the Supplemental Material. For $r$ close to one (weak selection) we find	
	\BE
	\phi = \frac{1-\widetilde r^{-1}}
				{1-\widetilde r^{-N}},
	\label{eq:Approx}
	\EE
with $\widetilde r = r + \langle1/k\rangle_c (1-r)$, and where $\langle1/k\rangle_c$ is the mean inverse degree among individuals who have at least one neighbour. This object depends on the connectivity $q$, which in turn depends on the interaction radius $R$. 
The weak-selection limit of the result for the complete graph [Eq.~\eqref{eq:CG_fixprob}] is recovered for $q=1$. If the interaction radius is small, and hence the interaction graph sparse ($q\rightarrow 0$), one finds $\langle 1/k\rangle_c\approx 1$, and $\widetilde r=1$, i.e., neutral selection. The finite connectivity of the dynamic interaction graph acts as a suppressor of selection.

The prediction of Eq.~\eqref{eq:Approx} is shown in the main panel of Fig. \ref{fig:PShear} (solid black line), and agrees with simulations for small Damk\"ohler numbers (fast flows). In the inset of the figure, we show the probability of fixation for different choices of the interaction radius $R$ for fast flows. In all cases, Eq.~(\ref{eq:Approx}) is seen to describe simulations well. The data demonstrates that, depending on the interaction radius, the fixation probability can take any value between the result for neutral selection and that predicted by the conventional well-mixed theory.

It is important to ask how fast the flow must be for Eq.~\eqref{eq:Approx} to be valid. Our approach requires that each individual experiences a newly sampled set of neighbours at each event, uncorrelated from its interaction partners at earlier events. Broadly speaking, our approach applies when the typical time $\tau$ between events involving a particular individual is larger than the mixing time $t_x$. The probability that any particular individual is involved in a given evolutionary event can be estimated as $1/N+(1-1/N)q$. This means that any individual typically participates in an event every $\tau=[\mbox{Da}(1+q(N-1))]^{-1}$ units of time. In the example of the parallel-shear flow $q\approx 0.07$. For a system with $N=100$ individuals, $\tau>t_x$ when $\mbox{Da}\lesssim 0.025$. If this condition is met we expect Eq.~(\ref{eq:Approx}) to apply. This is consistent with the data in Fig. \ref{fig:PShear}.

\section{Robustness and applicability to different flow fields}\label{sec:OtherFlows}

 In order to test our approach further we have simulated a range of different flows, as illustrated in Fig. \ref{fig:Table} and detailed further in the Supplemental Material. Similar to the parallel-shear flow, the double-gyre flow mixes the system at sufficiently small Damk\"ohler numbers (uniform entries in the connectivity matrix; $q_0,q_1\to q$). The chaotic blinking-vortex flow is approximately mixing. The non-chaotic blinking-vortex flow and the vortex-sink flow are not mixing, as can be seen in Fig. \ref{fig:Table}. The connectivity matrix resulting from these flows indicates clusters of particles which travel the system together; the set of neighbours of any one particle can remain correlated indefinitely.

\begin{figure*}[htb!!]
	\center
	\includegraphics[width=0.7\textwidth]{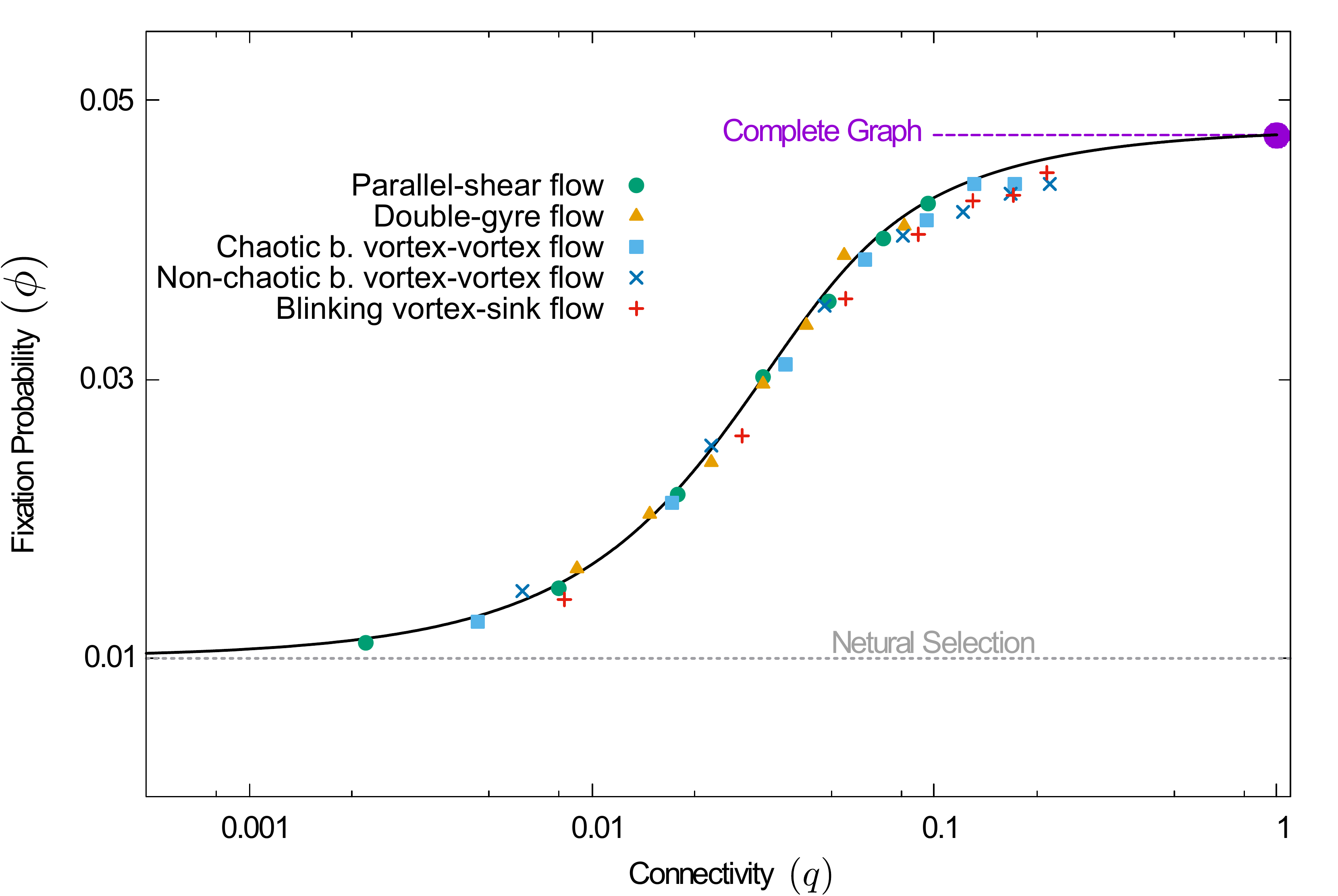}
	\caption{\textbf{Fixation probability as a function of connectivity.} 
				Varying the interaction radius interpolates between neutral selection and the theory based on complete graphs. The fast-flow theory applies throughout, provided the flow mixes the particles well. 
				The markers represent simulation results for different flows and different interaction radii, resulting in different connectivities, $q$. 
				Predictions of the fast-flow theory [Eq.~\eqref{eq:Approx}] are shown as the solid black line. 
				The conventional well-mixed theory [complete graph, Eq.~(\ref{eq:CG_fixprob})] is indicated by the filled circle at $q=1$. 
				The dashed gray line is for guidance only, and shows the result for neutral selection, $\phi=1/N$. 
				(Mutant fitness $r=1.05$, population size $N=100$).
				}
	\label{fig:AllFlowsQ}
	\end{figure*}

Results for the different flows are shown in Fig. \ref{fig:AllFlowsQ}. Different data points correspond to different choices of the interaction radius, resulting in different connectivities $q$. The conventional well-mixed theory is represented by the point $q=1$; our approach interpolates between this value and the one for neutral selection in the dilute limit $q\rightarrow0$. The data in the figure demonstrates that Eq.~(\ref{eq:Approx}) describes the fixation probability accurately for the flows that are mixing. Even for the non-chaotic blinking-vortex flow and the vortex-sink flow, it provides a good approximation.

\section{Dependence on the size of the population}
\begin{figure*}[t!!]
	\center
	\includegraphics[width=0.95\textwidth]{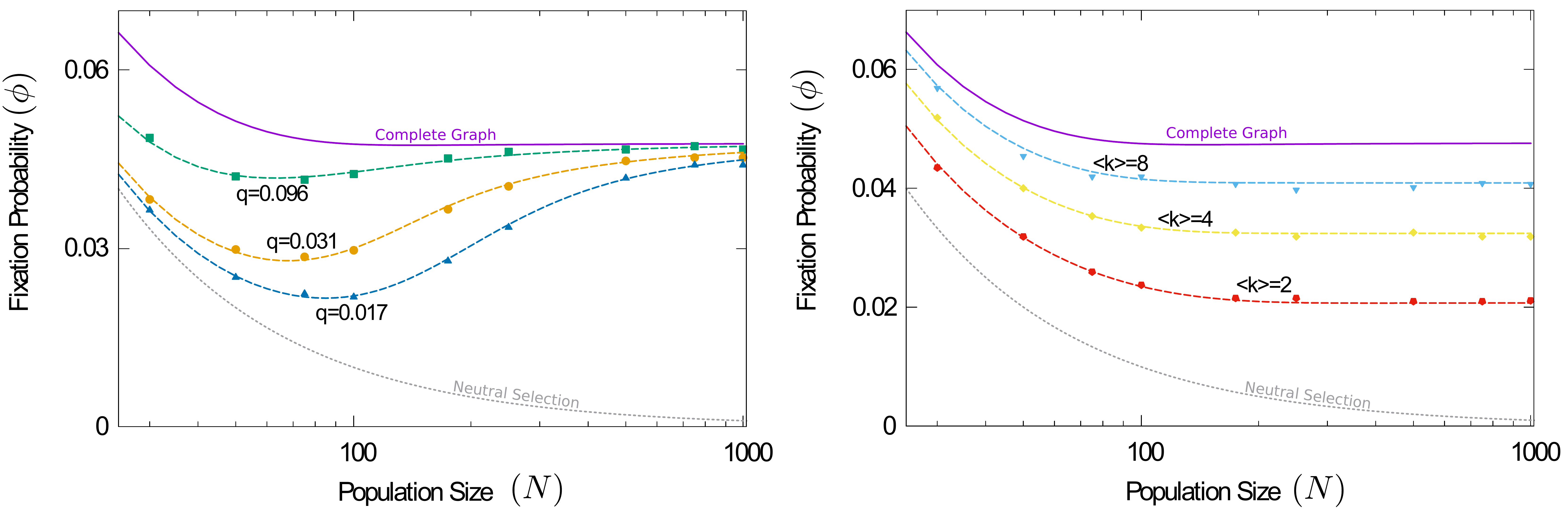}
	\caption{\textbf{Fixation probability as a function of population size.}  
				We show the fixation probability of a single mutant in a population of size $N$ for the parallel-shear flow. On the left-hand panel, the interaction radius $R$ is fixed as the population size is varied. This results in fixed connectivities, $q$, but the average number of neighbours of each particle increases with $N$. On the right-hand panel, the average number of neighbours, $\avg{k}$, was fixed by reducing the interaction radius  as the population size increases. Markers are from simulations. The conventional theory is shown as the thick purple line. Dashed coloured lines are the predictions of the fast-flow approach. The dashed gray line shows the result for neutral selection. ($r=1.05$, $\mbox{Da}=0.01$ in both panels.)}
	\label{fig:FixProbN}
	\end{figure*}

We show the fixation probability for different population sizes in Fig. \ref{fig:FixProbN}. 
In panel (a) we keep the connectivity $q$ fixed; the average number of neighbours of each individual is then $\avg{k}=(N-1) q$. Interestingly, this produces non-monotonic behaviour as a function of $N$, with minimal fixation probability at a certain population size. For small populations, the sampled neighbourhoods are so small that there is virtually no competition. The outcome of evolutionary events is dominated by the random composition of the set of neighbours of the removed individual, rather than by fitness. Effectively this results in neutral selection. In this regime, fixation of a single mutant becomes more difficult as $N$ increases, and the fixation probability $\phi$ is a decreasing function of $N$. For larger populations, neighbourhoods become large enough to provide a statistically more representative sample of the entire population. Selection becomes increasingly relevant, and the fixation probability of an advantageous mutation increases. In the limit of very large populations, the neighbourhoods are a statistically accurate sample of the entire population. Therefore, the traditional well-mixed theory, based on complete graphs, is recovered. We note that $\langle 1/k\rangle_c$ tends to zero in this limit, so that $\tilde r=r$, and the predictions of Eqs. (\ref{eq:CG_fixprob}) and $(\ref{eq:Approx})$ then agree. 
In panel (b) the average number of neighbours, $\avg{k}$, is kept fixed instead. Interactions are then always within local neighbourhoods, and the conventional complete-graph theory does not apply, even in large populations.

\section{Discussion}\label{sec:Conclusions}

In the existing literature, well-mixed populations are almost always associated with complete interaction graphs. Every member of the population is connected to every other member at all times. Competition and selection in an evolutionary event then takes place among all individuals. The term `well-mixed' suggests that these conditions can be achieved by stirring spatial systems. As we have shown, this is often not the case. Quantitative differences between the predictions of the conventional theory and simulations of stirred populations can be observed, even when the stirring is fast and when all pairs of individuals are equally likely to interact. We have presented an alternative approach, and demonstrated that our analytical description accurately predicts simulation results, even in situations where the conventional theory does not.

\begin{figure}[htb!!]
	\center
	\includegraphics[width=0.3\textwidth]{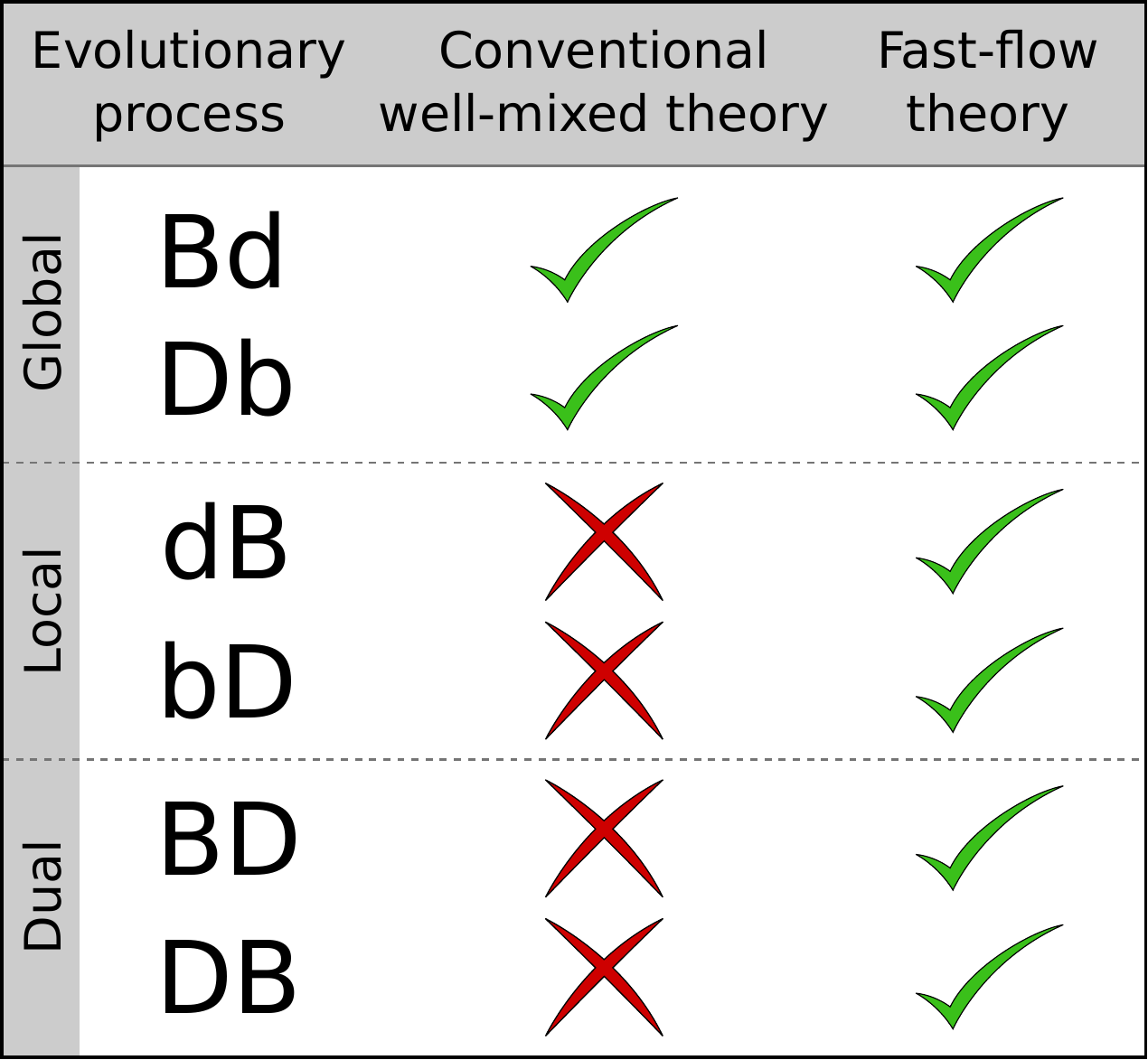}
	\caption{\textbf{Applicability of the  conventional well-mixed theory and the fast-flow theory.} 
				Evolutionary processes (described in the Supplemental Material) for which the predictions of each theory agree with simulation results are shown.
				Capital letters indicate selection in the birth or death step.
				In \textbf{Bd} and \textbf{Db} competition is global, and the conventional theory for well-mixed systems applies. The other four processes involve local selection and the conventional theory fails. 
				The fast-flow theory predicts simulation results in all cases.}
	\label{fig:mainprocesses}
	\end{figure}

So far we have only discussed one type of evolutionary dynamics, a death-birth process. In the model we have described, no competition takes place when the individual for removal is determined. The reproducing individual is selected from the neighbours of the removed and proportional to fitness. This is known as `local selection' and the process is referred to as a local death-birth process \cite{Kaveh2015,Hindersin2015,Fu2009}. Other variants are possible; for example, death-birth processes in which selection only acts when the individual for removal is chosen. This is known as a global death-birth process. Similarly selection can act at both the death and birth stages (death-birth process with dual selection). In very much the same way there are global and local birth-death processes, and birth-death processes with dual selection\cite{Kaveh2015,Zukewich2013}. We have tested the applicability of the conventional theory and of our approach to all six different types of processes (see Supplemental Material). We find that the conventional theory for well-mixed systems is accurate for processes in which selection only acts globally. A description based on a complete interaction graph is then appropriate. The conventional theory becomes invalid, however, when selection acts locally.  As summarised in Fig. \ref{fig:mainprocesses} the fast-flow approach we have developed applies for all six types of processes.
 
If the interaction range is limited, individuals do not compete against the entire population at any one time. Our analysis shows that this limited connectivity suppresses selection. Therefore, the conventional theory for well-mixed systems overestimates the fixation probability of advantageous mutants. Our results also suggest that a disadvantageous mutant is more likely to reach fixation when it has a small interaction radius. Similar results have recently been reported by Krieger et al. \cite{Krieger2017} for structured populations with mobile individuals.

The exact mechanics of evolution and the interaction range of individuals in biological or social systems are often difficult to determine. Mathematical modelling approaches frequently rely on well-mixed populations, due to the fact that these have analytical solutions. In some systems interaction may indeed occur over long distances, for example through signalling, chemical trails or the production of public goods \cite{VandeKoppel2015,Rauch2006,Menon2015,West2007}. Established approaches based on complete interaction graphs are then appropriate. Most systems, however, have a limited interaction range\cite{Tilman1997,Durrett1994}, and as consequence conventional well-mixed theories may not apply.

There is considerable theoretical work on the effects of local interactions in static structured populations \cite{Tilman1997,Durrett1994,Pattni2015,Nowak2010,Lieberman2005}.  Expressions for the fixation probability of invading mutants are known. However, populations in many social or biological systems are moving and the interaction network is dynamic. The fast-flow approach provides a tool that should prove useful for the modelling of situations in which individuals are in motion.


%


\section*{Acknowledgements}
FHA thanks Consejo Nacional de Ciencia y Tecnolog\'ia (CONACyT, Mexico) for support. VPM acknowledges support by Ministerio de Econom\'ia y Competitividad and Xunta de Galicia (MAT2015-71119-R, GPC2015/014), contributions by COST Action MP1305 and CRETUS Strategic Partnership (AGRUP2015/02).

\section*{Author contributions statement}
FHA carried out the simulations and analytical calculations. All authors contributed to designing the research, to analysing the data and to writing the paper.

\section*{Additional information}
We declare no competing interests.

\label{LastPageDoc}		

\clearpage
\clearpage
\setcounter{section}{0}		
\setcounter{page}{1}		
\setcounter{equation}{0}	
\setcounter{figure}{0}		
\renewcommand{\thesection}{S\arabic{section}} 		
\renewcommand{\thepage}{S\arabic{page}} 			
\renewcommand{\theequation}{S\arabic{equation}}  	
\renewcommand{\thefigure}{S\arabic{figure}}  		

\begin{center}
\vspace*{3cm}
{\LARGE \bf Fast flowing populations are not well mixed}\\
\vspace{2em}
{\large Francisco Herrer\'{i}as-Azcu\'{e}${}^{\rm 1}$, Vicente P\'{e}rez-Mu\~{n}uzuri${}^{\rm 2}$  and Tobias Galla${}^{\rm 1}$}\\

{\tt francisco.herreriasazcue@manchester.ac.uk, vicente.perez@cesga.es, tobias.galla@manchester.ac.uk}\\
\vspace{1em}
${}^{\rm 1}$Theoretical Physics, School of Physics and Astronomy, \\ The University of Manchester, Manchester M13 9PL, United Kingdom\\
\vspace{1em}

${}^{\rm 2}$Group of Nonlinear Physics, Faculty of Physics, \\ University of Santiago de Compostela E-15782 Santiago de Compostela, Spain
\vspace*{0.5cm}
\end{center}


\section{Evolutionary processes}\label{app:EvolProcesses}

Two types of processes are widely used in the literature to model evolutionary dynamics: \emph{birth-death} processes and \emph{death-birth} processes. A birth-death process in a spatial system or network occurs in two steps: first, an individual is chosen for reproduction from the entire population; then, one of its neighbours is chosen, and is replaced by an offspring of the first individual. In the second type of process, death-birth, the individual chosen in step one is removed from the population, and one of its neighbours is chosen to produce an offspring in the `vacant' place. (For clarity we stress again that two particles in the spatial system are considered to be `neighbours' if their distance is less than the interaction radius.)
\\

Each of these steps may or may not include an element of selection. Selection indicates that the choice of individual is based on fitness. Throughout our paper, we focus only on frequency-independent selection, i.e., the fitness of each type of individual only depends on the species it belongs to (mutant or wildtype), but not on the composition of its neighbourhood. In the literature, the most widely used update processes only include selection in the reproduction step. In principle, however, each individual can carry two types of fitness \cite{Kaveh2015}. One is reproductive fitness; when competition occurs in the reproduction stage, it determines the probability that an individual is picked. If selection occurs during the choice of an individual for removal, we think of the resilience (against removal) as a `death fitness'. It plays the same role as the reproduction fitness, but in the removal stage. Individuals with a higher  `death fitness' are less likely to be chosen for removal. Without loss of generality, we set both fitnesses of the wildtype to one. We write $r$ for the mutant's reproductive fitness, and $1/d$ for its death fitness. The quantity $d$ then describes a propensity to die.
\\

Six possible processes can then be considered. The first three are birth-death processes. Selection can occur only in the birth step (\textbf{Bd}), only in the removal step (\textbf{bD}), or in both (\textbf{BD}). Similarly, for death-birth processes one has \textbf{Db}, \textbf{dB} and \textbf{DB}. Capital letters indicate that the step involves selection, and lower case letters indicate the absence of selection.

It is important to stress that the first individual in each birth-death or death-birth event is chosen from the {\em entire} population. If selection occurs at this step, this selection is {\em global.} The second individual is chosen among the neighbours of the first. Therefore, if selection occurs in this phase, it is {\em local} selection.

The most general birth-death process involves selection in both steps (\textbf{BD}, $r\neq 1$, $d\neq1$). If $d=1$, the birth-death process is of the type \textbf{Bd}. Selection takes place when individuals are chosen from the entire population, and it is hence global selection. For this reason, the process is also referred to as a global birth-death process. A dynamics of the type \textbf{bD} is a local birth-death process process ($r=1$). The nomenclature for global and local death-birth processes is similar (\textbf{Db} denotes the global death-birth process, and \textbf{dB} the local death-birth process, respectively).

\section{Fixation probabilities in the limit of fast flows}\label{app:Theory}
\subsection{Setup and notation}
A focal individual is chosen in step one of a birth-death or death-birth process. Interaction then occurs with one of its neighbours (particles within its interaction radius $R$). In the limit of fast flow, we assume that this neighbourhood is sampled from the entire population at random (excluding the focal individual). Each individual is in the neighbourhood with probability $q$, and it is not a neighbour with probability $1-q$.

The connectivity $q$ will depend on the interaction radius. The parallel-shear flow, for example, generates a uniform stationary density of particles. Therefore, any randomly chosen individual will be a neighbour of the focal individual with probability $q=\pi R^2/A$, where $A$ is the total area of the system. This assumes $0<R<(A\pi)^{-1/2}$ and periodic boundary conditions. For $R\ge(A\pi)^{-1/2}$ the network is fully connected ($q=1$).  

With these assumptions the degree distribution is binomial; the probability that the focal individual has exactly $k$ neighbours is
	\be
	P_{k}=\binom{N-1}{k}q^{k}\left(1-q\right)^{N-1-k}.
	\label{eq:Wellmixed_probk}
	\ee

Assume now that there are $m$ mutants in the population and $N-m$ wildtypes. If the focal individual is a mutant, the probability that there are $l$ mutants among its $k$ neighbours is 
	\be
	P^{(m)}(l | k) =\binom{k}{l}\left(\frac{m-1}{N-1}\right)^{l}\left(1-\frac{m-1}{N-1}\right)^{k-l}.
	\ee
Instead, if the focal individual is a wildtype the probability that $l$ of its $k$ neighbours are mutants is
	\be
	P^{(w)}(l | k)=\binom{k}{l}\left(\frac{m}{N-1}\right)^{l}\left(1-\frac{m}{N-1}\right)^{k-l}.
	\ee

We next consider the situation in which $k$ individuals compete (for example to replace a removed individual). Assuming that $l$ of the $k$ individuals are mutants (with fitness $r$) and $k-l$ are wildtypes (with fitness one) we write $g_{kl}(r)$ for the probability that a mutant is selected. We have
	\be
	g_{kl}(r)=\frac{l r}{(k-l)+l r}.
	\ee
The quantity $1-g_{kl}(r)$ is the probability that a wildtype is selected in this situation. If selections acts during the removal step, the probability that the individual chosen for removal is a mutant is $g_{kl}(d)$.

It is also useful to define
	\BE
	H^{(m)}(r)&=&\sum_k\sum_{l=0}^k P_k\, P^{(m)}(l | k) [1-g_{kl}(r)],	\nn \\
	H^{(w)}(r)&=&\sum_k\sum_{l=0}^k P_k\, P^{(w)}(l | k) g_{kl}(r),			
	\label{eq:H}
	\EE
where the double sums run over all neighbourhood compositions of the focal individual. The quantity $H^{(m)}$ describes the probability with which a wildtype is selected among the neighbours of a mutant focal individual, and $H^{(w)}$ the probability that a mutant is selected from the neighbourhood of a wildtype.
\\

In each death-birth or birth-death event, the number of mutants in the population may increase by one ($m\to m+1$), decrease by one ($m\to m-1$), or remain unchanged. We write $T_{m}^{+}$ for the probability that $m$ increases by one, and $T_{m}^{-}$ for the probability that $m$ decreases by one. For birth-death processes we have
	\BE
	T_{m}^{+} & = & \frac{mr}{mr+N-m}  \times H^{(m)}(d),			\nn	\\
	T_{m}^{-} & = & \frac{N-m}{mr+N-m} \times H^{(w)}(d),
	\EE
where the fraction on the right-hand side in the expression for $T_m^+$ is the probability that the individual selected for birth is a mutant, or that it is a wildtype in the expression for $T_m^-$.

For death-birth processes we have
	\BE
	T_{m}^{+}&=&\frac{N-m}{md+N-m}\times H^{(w)}(r),				\nn	\\
	T_{m}^{-}&=&\frac{md}{md+N-m}\times H^{(m)}(r).
	\EE

\subsection{Fixation probability}

For any one-step process, the probability of fixation of a single mutant in a population of size $N$ can be written as \cite{Traulsen2009}  
	\BE
	\phi=\frac{1}{1+\sum\limits _{j=1}^{N-1}\prod\limits _{m=1}^{j}\gamma_m},
	\label{eq:FixProb}
	\EE
where $\gamma_m=T_m^-/T_m^+$, sometimes known as the evolutionary drift.
\\

For the \textbf{BD} and \textbf{DB} processes we have, respectively,
	\BE
	\gamma^{BD}_m(r,d) & = & \frac{N-m}{mr}\frac{H^{(w)}(d)}{H^{(m)}(d)},\nonumber \\
	\gamma^{DB}_m(r,d) & = & \frac{md}{N-m}\frac{H^{(m)}(r)}{H^{(w)}(r)}.
	\label{eq:Wellmixed_drift}
	\EE
We note that $\gamma^{BD}_m(r,d)=\left[\gamma^{DB}_m(d,r)\right]^{-1}$.

Substituting Eqs. (\ref{eq:Wellmixed_drift}) into Eq.~(\ref{eq:FixProb}) yields the fixation probability for the two types of processes:
	\BE
	\phi_{BD}\left(r,d\right) & = & \frac{1}{1+\sum\limits _{j=1}^{N-1}\prod\limits _{m=1}^{j}
		\frac{N-m}{mr}\frac{H^{(w)}(d)}{H^{(m)}(d)}}	\label{eq:FF_fixprobBD}		\\				 
	\phi_{DB}\left(d,r\right) & = & \frac{1}{1+\sum\limits _{j=1}^{N-1}\prod\limits _{m=1}^{j}
		\frac{md}{N-m}\frac{H^{(m)}(r)}{H^{(w)}(r)}}	\label{eq:FF_fixprobDB}
	\EE
These closed-form expressions can readily be evaluated numerically. 
\\

For the global processes, there is no selection in the second step of evolutionary events, and so the expressions for $H^{(m)}$ and $H^{(w)}$ simplify considerably. If followed through, Eqs. \eqref{eq:FF_fixprobBD} and \eqref{eq:FF_fixprobDB} can be reduced to  	
\BE
	\phi_{BD} (r,d=1)&=&\phi_{Bd} \equiv \frac{1-r^{-1}}{1-r^{-N}},	\nn \\
	\phi_{DB} (r=1,d)&=& \phi_{Db} \equiv  \frac{1-d}{1-d^{N}},
	\label{eq:FF_fixprob_global}
	\EE
which are the well-well known results for the complete graph \cite{Traulsen2009}. It is important to note that these equations are for processes in which selection acts in the first stage of the birth-death or death-birth events. This means that selection acts globally, and that \emph{all} $N$ individuals in the population are in competition. The \textbf{dB} process used in the main text is slightly different. In that model, selection does not act in the first stage of each event, it acts in the second. Therefore, unless self-replacement is allowed, selection is made among $N-1$ individuals  even if the interaction graph is complete (the entire population, except the individual chosen in the first stage). This leads to a slightly different expression, see e.g. Eq.~\eqref{eq:CG_fixprob} of the main text, even though both processes have selection only in the birth stage. As the system size increases, however, the difference between \textbf{dB} and \textbf{Bd} on a complete graph becomes small, and so the result in Eq.~\eqref{eq:CG_fixprob} tends to that in the first equation of (\ref{eq:FF_fixprob_global}).
\\

The predictions for the local birth-death and death-birth processes (\textbf{bD} and \textbf{dB}) do not reduce to expressions as simple as those in Eq.~(\ref{eq:FF_fixprob_global}). This is perhaps to be expected from previous studies of local processes on static networks, which have shown that the traditional well-mixed theory is only valid for a very narrow set of graphs \cite{Pattni2015}. 

\subsection{Approximation in the limit of weak selection}
 We now proceed to simplify the expressions in Eqs. \eqref{eq:FF_fixprobBD} and \eqref{eq:FF_fixprobDB}. We will focus on the case of the \textbf{DB} process, but similar expressions can be obtained for \textbf{BD} upon replacing $d$ by $r$. We begin by simplifying the expressions for $H^{(m)}$ and $H^{(w)}$ in Eq.~\eqref{eq:H}, 
 \BE
	H^{(m)}(d)& = & 
		\sum\limits _{k=1}^{N-1}P_{k}\sum\limits _{l=0}^{k}\binom{k}{l}\left(\frac{m-1}
			{N-1}\right)^{l}\left(1-\frac{m-1}{N-1}\right)^{k-l}\left(\frac{k-l}{ld+k-l}\right),		\nn \\
	H^{(w)}(d)  & = & 
		\sum\limits _{k=1}^{N-1}P_{k}\sum\limits _{l=0}^{k}\binom{k}{l}\left(\frac{m}{N-1}\right)^{l}
			\left(1-\frac{m}{N-1}\right)^{k-l}\left(\frac{ld}{ld+k-l}\right).		\label{eq:binom}
	\EE
Since $\frac{k-l}{ld+k-l}=1-\frac{ld}{ld+k-l}$,  we only need to compute objects of the type $\sum\nolimits _{l=0}^{k}\binom{k}{l}x^l(1-x)^{k-l}\left(\frac{ld}{ld+k-l}\right)$. We expand in powers of $1-d$, and keeping only terms to first order we obtain (after re-arranging terms)
	\BE
	\frac{ld}{ld+k-l} &=&  1+\left(\frac{l}{k}-1\right)\sum_{i=0}^{\infty}\left(\frac{l}{k}
							\left(1-d\right)\right)^{i} 										\nn \\
					  &=&\frac{l}{k}d+\left(\frac{l}{k}\right)^2(1-d) 
					  		+\mathcal{O}\left(\left(1-d\right)^{2}\right).
	\EE

Using this approximation we find
	\BE
	\sum\limits _{l=0}^{k}\binom{k}{l}x^{l}\left(1-x\right)^{k-l}\left(\frac{ld}{ld+k-l}\right) 
		& =& \sum\limits _{l=0}^{k}\binom{k}{l}x^{l}\left(1-x\right)^{k-l}
		\left[d\frac{l}{k}+\left(\frac{l}{k}\right)^{2}\left(1-d\right)\right]	+
		\mathcal{O}\left(\left(1-d\right)^{2}\right)											\nn \\
	 & = & d~x+\left(1-d\right)\left(\frac{x}{k}+\frac{x^{2}\left(k-1\right)}{k}\right)+
	 	\mathcal{O}\left(\left(1-d\right)^{2}\right)											\nn \\
	 & = & d~x+\left(1-d\right)x^{2}+\frac{\left(1-d\right)\left(x-x^{2}\right)}{k}+
	 	\mathcal{O}\left(\left(1-d\right)^{2}\right).
	\EE
The expressions in Eq.~(\ref{eq:binom}) can therefore be approximated as
	\BE
	H^{(m)}(d) & \approx & \sum\limits _{k=1}^{N-1}P_{k}
		\left[1-d\left(\frac{m-1}{N-1}\right)-\left(1-d\right)\left(\frac{m-1}{N-1}\right)^{2}
		-\frac{\left(1-d\right)\left(\left(\frac{m-1}{N-1}\right)-\left(\frac{m-1}{N-1}\right)^{2}
		\right)}{k}+\dots\right]																			\nn \\
	 & = & \left(\frac{N-m}{N-1}\right)\left[\left(\sum\limits _{k=1}^{N-1}P_{k}\right)+
	 	\left(1-d\right)\left(\frac{m-1}{N-1}\right)\left(\sum\limits _{k=1}^{N-1}
	 	P_{k}\frac{k-1}{k}\right)\right]	+ \mathcal{O}\left(\left(1-d\right)^{2}\right)														\nn \\
	 & = & \left(\frac{N-m}{N-1}\right)q_c \left[1+\left(1-d\right)\left(\frac{m-1}{N-1}\right)
	 	\left(1-\avg{\frac{1}{k}}_c\right)\right]+\mathcal{O}\left(\left(1-d\right)^{2}\right),													\nn	\\
	H^{(w)}(d) & \approx & \sum\limits _{k=1}^{N-1}P_{k}
		\left[d\left(\frac{m}{N-1}\right)+\left(1-d\right)\left(\frac{m}{N-1}\right)^{2}+
		\frac{\left(1-d\right)\left(\left(\frac{m}{N-1}\right)-\left(\frac{m}{N-1}\right)^{2}
		\right)}{k}+\dots\right]																			\nn \\
	 & = & \left(\frac{m}{N-1}\right)\left[\left(\sum\limits _{k=1}^{N-1}P_{k}\right)-
	 	\left(1-d\right)\left(\frac{N-m-1}{N-1}\right)\left(\sum\limits _{k=1}^{N-1}P_{k}
	 	\frac{k-1}{k}\right)\right]+\mathcal{O}\left(\left(1-d\right)^{2}\right)																	\nn	\\
	 & = & \left(\frac{m}{N-1}\right)q_c \left[1+\left(1-d\right)\left(\frac{N-m-1}{N-1}\right)
	 	\left(1-\avg{\frac{1}{k}}_c\right)\right]+\mathcal{O}\left(\left(1-d\right)^{2}\right).
	\label{eq:Wellmixed_Neighbourhood_operators_approx}
	\EE
In these expressions we have written
	\be
	q_c\equiv \sum_{k=1}^{N-1} P_k=1-P_0
	\ee
for the probability that a randomly chosen individual has at least one neighbour, and
	\be
	\avg{\frac{1}{k}}_c \equiv \frac{1}{q_c}\sum_{k=1}^{N-1} \frac{P_k}{k}.
	\ee
This expression describes the average inverse degree of all nodes with at least one neighbour.

Substituting Eqs. \eqref{eq:Wellmixed_Neighbourhood_operators_approx} into Eqs. \eqref{eq:Wellmixed_drift}, and again expanding in powers of $d-1$ and $r-1$, respectively, yields
	\BE
	\gamma^{BD}\left(r,d\right) & = & \frac{1}{r}\bigg[1-\left(1-d\right)\left(1-\avg{1/k}_c\right)\bigg]
										+\mathcal{O}\left(\left(1-d\right)^{2}\right),			\nn \\
	\gamma^{DB}\left(d,r\right) & = & d\left[\frac{1}{1-\left(1-r\right)\left(1-\avg{1/k}_c\right)}\right]
										+\mathcal{O}\left(\left(1-r\right)^{2}\right).
	\label{eq:Wellmixed_drift_approx_fixedk_Ninf}
	\EE
In contrast with Eqs. \eqref{eq:Wellmixed_drift}, the first-order expressions on the right-hand side are now independent of $m$. It is therefore straightforward to evaluate the general expression for the fixation probability of a single mutant [Eq.~\eqref{eq:FixProb}]. For the \textbf{BD} process this leads to
	\BE
	\phi_{BD}\left(r,d\right) & = & \frac{1}{1+\sum\limits _{j=1}^{N-1}
		\prod\limits _{m=1}^{j}\frac{1}{r}\bigg[1-\left(1-d\right)\left(1-\avg{1/k}_c\right)\bigg]} \nn \\
		&=&\frac{1-\left(\frac{d+\avg{1/k}_c(1-d)}{r}\right)}{1-\left(\frac{d+\avg{1/k}_c(1-d)}{r}\right)^N},
	\label{eq:approxBD}
	\EE
where we have neglected contributions of order $(1-d)^2$ and higher.

In the case of the global process (\textbf{Bd}, $d=1$), this reduces to Eq.~\eqref{eq:FF_fixprob_global}. For the local process (\textbf{bD}) we have $r=1$, and so Eq.~(\ref{eq:approxBD}) reduces to
	\be
	\phi_{BD}\left(r=1,d\right) = \phi_{bD}=\frac{1-\widetilde d}{1-\widetilde d^N},
	\label{eq:approxlocbD}
	\ee
with 
	\be
	\widetilde d = d+\avg{1/k}_c(1-d).
	\ee

Similarly, for the \textbf{DB} process we have (disregarding corrections of order $(1-r)^2$ and higher)
	\BE
	\phi_{DB}\left(d,r\right) & = & \frac{1}{1+\sum\limits _{j=1}^{N-1}
		\prod\limits _{m=1}^{j}d\left(\frac{1}{1-\left(1-r\right)\left(1-\avg{1/k}_c\right)}\right)} \nn \\
		&=&\frac{1-\left(\frac{d}{r+\avg{1/k}_c(1-r)}\right)}{1-\left(\frac{d}{r+\avg{1/k}_c(1-r)}\right)^{N}}.
	\label{eq:approxDB}
	\EE
Setting $r=1$ recovers the result for the global death-birth process (\textbf{Db}) in Eq.~\eqref{eq:FF_fixprob_global}. For the local process (\textbf{dB}) we have $d=1$, and so Eq.~(\ref{eq:approxDB}) reduces to 
	\be
	\phi_{DB}\left(1,r\right) = \phi_{dB} = \frac{1-\widetilde r^{-1}}{1-\widetilde r^{-N}},
		\label{eq:approxlocdB}
	\ee
with
	\be
	\widetilde r= r+\avg{1/k}_c(1-r).
	\ee
This corresponds to the local death-birth process (\textbf{dB}), and is the focus of the main text (see Eq.~\eqref{eq:Approx}).

\bigskip

\begin{figure*}[htb!!]
	\center
	\includegraphics[width=0.95\textwidth]{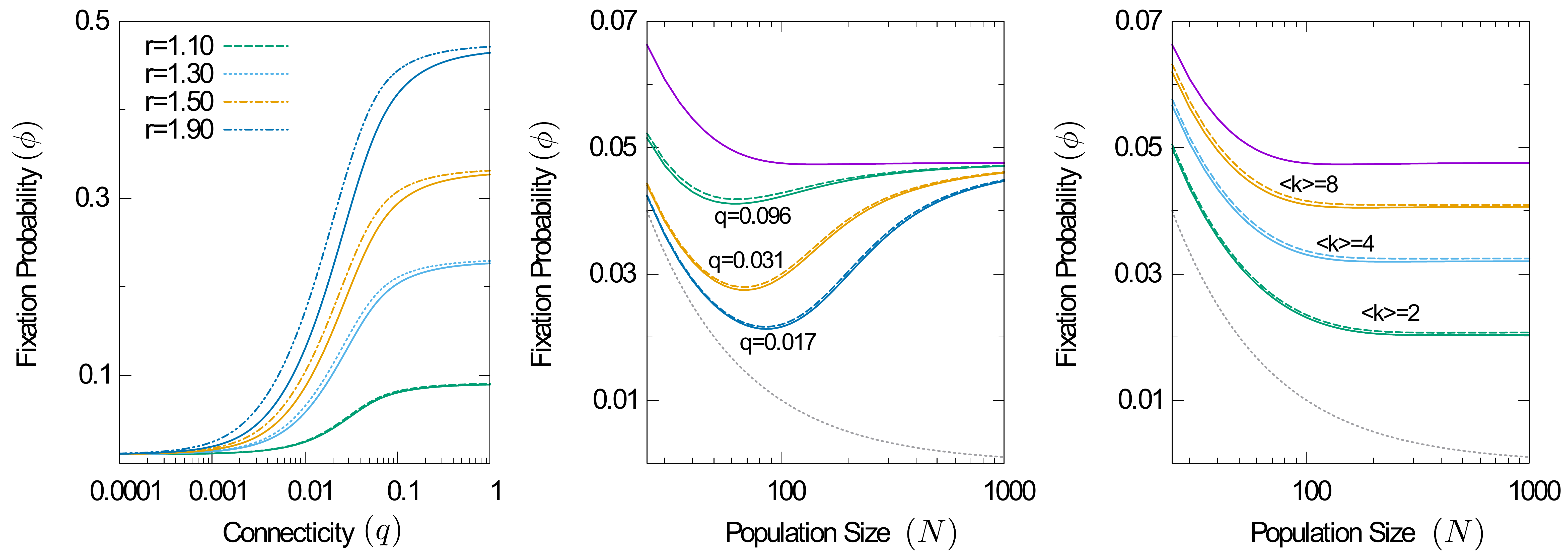}
	\caption{\textbf{Test of the weak-selection approximation.} 
				The left panel shows the fixation probability as a function of the connectivity $q$ for different mutant fitnesses. 
				The central and right panel show the fixation probability of a mutant with fitness $r=1.05$ as a function of the population size ($q$ fixed in central panel, average degree $\avg{k}$ fixed on the right).
				Continuous lines show the equations prior to the approximation [Eq.~\eqref{eq:FF_fixprobDB}], and dashed lines show the equations after the approximation [Eq.~\eqref{eq:approxlocdB}]. As can be seen, the approximation is valid for all system sizes, but is sensitive at large fitnesses. 
				}
	\label{fig:Approx}
	\end{figure*}

To test the accuracy of the weak-selection approximation leading to Eq.~\eqref{eq:approxlocdB} we compare its predictions against that of the full expression of Eq.~\eqref{eq:FF_fixprobDB} in Fig. \ref{fig:Approx}. In the left-most pane, we plot the fixation probability as a function of the connectivity for different mutant fitnesses. Since the approximation requires weak-selection, deviations are expected when $r$ is significantly different from one. In the other two panels, we show the fixation probability as a function of the population size (see also Fig. \ref{fig:FixProbN}), keeping the connectivity $q$ constant (central panel), or fixing the average number of neighbours $\avg{k}$ instead (panel on the right). As can be seen in the figure, the approximation remains valid for any system size.

\subsection{Validity of well-mixed and fast-flow theories for the different evolutionary processes}
In this section we comment on the applicability of the conventional well-mixed theory and that of our approach to the six types of birth-death and death-birth processes briefly discussed in the main text. Two individuals directly participate in each evolutionary event. The first is chosen from the entire population, and the second from the neighbours of the first individual. Two particles are `neighbours' when the distance between them is at most $R$ (the interaction radius). The ordering (birth-death versus death-birth) indicates whether the individual that is chosen first is removed from the population (death-birth) or whether it reproduces (birth-death). Capital letters in the acronyms indicate whether selection takes place in each of the two steps, i.e., in \textbf{BD} and \textbf{DB} processes, both steps involve selection, in \textbf{Db} and \textbf{Bd} only the first step (global selection), and in \textbf{dB} and \textbf{bD} only the second step (local selection).

\subsubsection{Global processes: Db and Bd}
The global processes are obtained by setting $r=1$ in a general death-birth process (resulting in \textbf{Db}), or setting $d=1$ in a general birth-death process (resulting in \textbf{Bd}). Our approach for the fast-flow limit then reduces to the conventional well-mixed theory, and the predictions for the fixation probability of a single mutant are those in Eqs. \eqref{eq:FF_fixprob_global}. These agree well with simulations, as shown in Fig. \ref{fig:Processes} (compare crosses and dark purple line). For simplicity we use $d=1/r$; in this case the two equations in \eqref{eq:FF_fixprob_global} are identical.
 
\subsubsection{Local processes: dB and bD}
The local processes are obtained by setting $d=1$ in a general death-birth process (resulting in \textbf{dB}), or setting $r=1$ in a general birth-death process (resulting in \textbf{bD}). For weak selection, the predictions from our theoretical approach for the fixation probability of a single mutant is then given by Eqs. (\ref{eq:approxlocbD}) for \textbf{bD}, and Eq.~(\ref{eq:approxlocdB}) for \textbf{dB}. These are different from the predictions of the conventional theory for well-mixed systems, Eqs. (\ref{eq:FF_fixprob_global}). The simulation data (empty squares and circles) in Fig. \ref{fig:Processes} agree with the predictions of Eqs. (\ref{eq:approxlocbD}, \ref{eq:approxlocdB}), shown as a dashed blue line (for $d=1/r$ the predictions of these two equations are indistinguishable on the scale of the graph). The conventional theory (dark purple line) deviates from the simulation data.
 
\subsubsection{Selection in both steps: DB and BD}
\begin{figure*}[t!!]
	\center
	\includegraphics[width=0.95\textwidth]{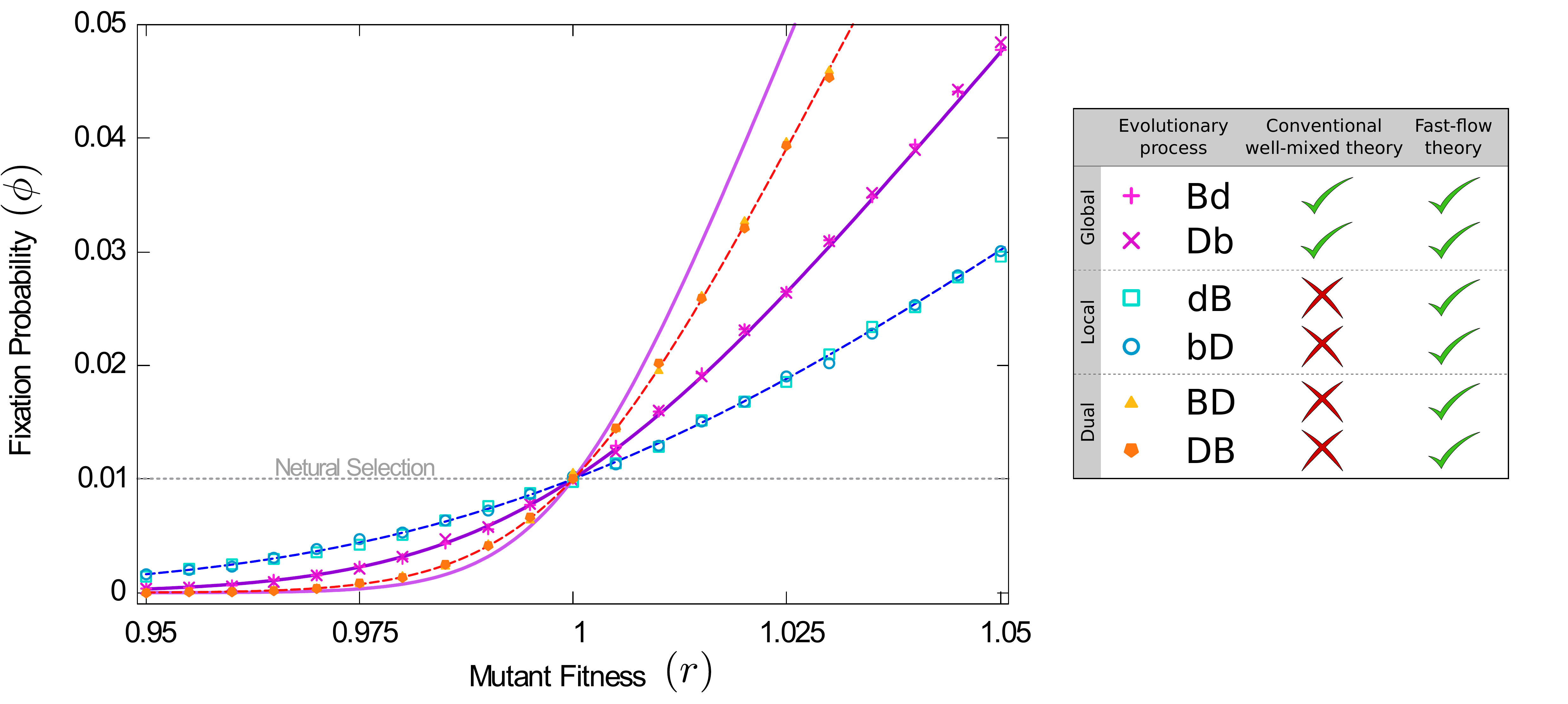}
	\caption{\textbf{Fixation probability as a function of fitness for the different update processes.} 
				Continuous thick lines show the conventional well-mixed theory for processes with dual selection (light purple) or selection in only one step (dark purple). 
				Dashed lines show the fast-flow theory for dual selection (red), or local selection (blue). We use $d=1/r$. 	The theoretical predictions for \textbf{BD} and \textbf{DB} are then indistinguishable on the scale of the figure, and similarly for the pairs \textbf{Bd-Db}, and \textbf{bD-dB} respectively. Simulation results are for the parallel-shear flow, with $\Da=0.01$, $R=0.1$ and $N=100$. 
			}
	\label{fig:Processes}
	\end{figure*}

In the \textbf{DB} and \textbf{BD} processes selection takes place in both steps. The predictions of our approach are those of Eqs. (\ref{eq:approxBD}) and (\ref{eq:approxDB}) and are shown as a dashed red line in Fig. \ref{fig:Processes} (for $d=1/r$ the predictions of the two equations are again indistinguishable on the scale of the figre). They agree with the simulation results (solid triangles and pentagons). The predictions of the complete-graph theory for the processes with dual selection is also shown for comparison (light purple line).

\section{Description of the flows}\label{app:Flows}
To test our analytic predictions we have used a selection of different flows, as summarised in Fig. \ref{fig:Table}. These are all planar flows and, as a consequence, an explicit time dependence is necessary to allow for chaotic motion. Each flow is defined by a flow field, $v_x(x,y,t)$ and $v_y(x,y,t)$. We treat the individuals in the population as Lagrangian particles; their motion is governed by $\dot x = v_x(x,y,t),~ \dot y=v_y(x,y,t)$. We write $\bv=(v_x,v_y)$

All flows we have used are periodic $\bv(x,y,t+T)=\bv(x,y,t)$, where $T$ is the period (we use $T=1$ throughout). The only exception is the parallel-shear flow, which additionally involves a phase, randomly chosen at the beginning of each half period. Details of the flows can be found in the literature \cite{Ottino1989,Kundu1990,Acheson1990,Young2010,Neufeld2010}. We briefly describe their main features below. 

\subsection{Parallel-Shear}

In this flow, particles move in the domain $0\leq x,y\leq 1$, with periodic boundary conditions. 

For the first half of the period ($nT\le t<nT+T/2$), the particles move horizontally, with a velocity which depends on their vertical position. Specifically
	\BE
	v_x(x,y,t)&=& V_{\rm max}\sin\left(2\pi y_{t}+\phi\right), \nonumber \\
	v_y(x,y,t)&=&0.
	\EE
The constant pre-factor $V_{\rm max}$ sets the overall magnitude of the flow. We use $V_{\rm max}=1.4$. The phase $\phi$ is drawn randomly from the interval $[0,2\pi)$ at the beginning of each half period.

During the second half of each period [$nT+T/2\le t<(n+1)T$], the particles move vertically, with a velocity that depends on their horizontal position,
	\BE
	v_x(x,y,t)&=&0,\nonumber \\
	v_y(x,y,t)&=& V_{\rm max}\sin\left(2\pi x_{t}+\phi\right).
	\EE

Within each half period the velocity of each particle is constant. In the first half of each period numerical integration can therefore be carried out using
	\be
	x(t+\Delta t)=x(t)+\Delta t\,V_{\rm max}\sin\left[2\pi y_{t}+\phi\right],
	\ee
and in a similar way for the second half of each period.

The time step $\Delta t$ does not need to be kept small, so long as the end of the half-period is not reached. In practice it is convenient to first schedule the times at which evolutionary events occur (they occur at fixed intervals, calculated based on the Damk\"ohler number). At any one time, the time step $\Delta t$ can then be chosen as the remaining time until the next evolutionary event or the end of the next half-period (whichever is shorter). This generates a very efficient numerical integration scheme.

\subsection{Double-Gyre}

The spatial domain is now given by $0\leq x\leq 2$ and $0\leq y\leq 1$. No particular boundary conditions apply, as particles cannot leave the domain.

A \emph{clock-wise} rotating gyre is centred on $\left(0.5,0.5\right)$, and a \emph{counter-clock-wise} rotating gyre on $\left(1.5,0.5\right)$. Each gyre rotates the particles in a spiral motion. In the  absence of further external driving there is no flow of particles across the line $x=1$. However, if the transport barrier between the two gyres is driven back and forth, particles can move between the two spirals. 

The flow field is given by
	\BE
	v_x(x,y,t)&=&\pi M\sin\left[\pi a(t)x^{2}+\pi b(t)x\right]
			\cos\left(\pi y\right),		\nn \\
	v_y(x,y,t)&=&\pi M\left[2a(t)x+b(t)\right]
			\cos\left[\pi a(t)x^{2}+\pi b(t)x\right]\sin\left(\pi y\right),
	\EE
where		
	\BE
		a(t) & = & u_{0}\sin\frac{2\pi t}{T},		\nn \\
		b(t) & = & 1-2u_{0}\sin\frac{2\pi t}{T}.	
	\EE
	
The parameter $M$ sets the amplitude of the flow around the gyres, and $u_{0}$ controls the movement of the barrier between the two gyres. We use $M=1.4$ and $u_{0}=0.4$. 

We simulate the flow using an Euler scheme, with time step $\Delta t =0.001$.

\subsection{Blinking Vortex-Vortex}

The blinking vortex-vortex flow describes motion around two vortices which are `active' during alternating times. For the first half of the period the particles rotate around a vortex located at $x_c=-b$, and for the second half of the period the centre of rotation takes the position $x_c=b$. In the simulations we use $b=0.25$.

We first describe the motion of particles around a vortex with fixed centre at $(x_c,0)$. In this case the equations of motion are\cite{Neufeld2010}
	\BE
	v_x(x,y)=-\frac{\Gamma y}{\tilde x+y^2} \nonumber \\
	v_y(x,y)=\frac{\Gamma \tilde x}{\tilde x^2+y^2},
	\EE
where $\tilde{x}(t) = x(t) - x_c$. This motion keeps the distance from the centre of the vortex constant ($r^2\equiv\tilde x^2+y^2=\mbox{const}$), and results in an angular velocity $\omega=\Gamma/r^2$ which is constant for each particle, but which depends on the distance from the vortex centre. The tangential component of the velocity $\bv$ is proportional to $1/r$. The constant $\Gamma$ sets the scale of the flow velocity.

During each half-period (i.e., during rotations about a fixed vortex) the radial distance from the vortex centre remains constant. The motion is implemented conveniently through application of a rotation matrix,
	\BE
	\left(\begin{array}{c} \tilde{x}(t+\Delta t) \\ y(t+\Delta t)\end{array}\right)=\left(\begin{array}{cc} \cos(\Delta\theta) & -\sin(\Delta\theta) \\ \sin(\Delta\theta) & \cos(\Delta\theta) \end{array}\right) \left(\begin{array}{c} \tilde{x}(t) \\ y(t)\end{array}\right),
	\EE
where
	\be
		\Delta \theta  = \frac{\Gamma}{\tilde{x}^{2}(t)+y^{2}(t)}\Delta t,
	\ee
is the rotation angle in time the interval $\Delta t$. As in the parallel-shear flow, there is no requirement to use a small time step $\Delta t$; it can be chosen as the time until the end of the next half-period or the time until the next evolutionary event (whichever comes sooner).	
\\
	 
Depending on the choice of parameters and the initial conditions, the blinking vortex-vortex flow can either be chaotic or non-chaotic:
\\

\paragraph{Chaotic}
For $\Gamma\gtrsim\Gamma_c=0.14$ the flow is chaotic\cite{Ottino1989}. The simulations for the chaotic case shown in the main text correspond to $\Gamma\approx0.2$.

\paragraph{Non-Chaotic}
If $\Gamma$ is below the critical value, the flow is not chaotic. Instead domains separated by transport barriers are obtained\cite{Neufeld2010}. The flow can then not be expected to be well mixing. Simulations shown in the main text correspond to $\Gamma\approx0.01$.

Initially we place all particles in the region $-0.24\leq x,y\leq0.24$. The domain in which the particles move is in principle not bounded. However, at long times all particles in our simulations are found within a fixed bounded area.

\subsection{Blinking Vortex-Sink}

This flow is very similar to the blinking vortex-vortex flow, but one of the vortices is replaced by a sink. When the sink is active particles are attracted towards its centre, and there is no angular motion. Specifically, in polar coordinates this is of the form 
\BE
\dot{\theta}&=&0 ,\nonumber \\
\dot{r}&=&\frac{m}{2\pi r},
\EE
 where $r$ is the distance from the sink. This translates into $\frac{d}{dt} r^2 = m/\pi$.

The position of the particles is updated in the same way as in the blinking vortex-vortex flow for the first half of the period. When the sink is active (second half of each period), the numerical scheme involves resizing the radial distance from the sink by a factor $\lambda$. We write $\tilde{x}(t) = x(t) - x_s$, where $x_s$ is the location of the sink. We then have 
	\BE
	\tilde{x}(t+\Delta t) &=& \lambda \tilde{x}(t) \nonumber \\
	y(t+\Delta t)&=&\lambda y(t).
	\EE
The scale factor is
	\be
	\lambda=\sqrt{1-\frac{\Delta t\,m}{\pi\left(\tilde{x}^{2}(t)+y^{2}(t)\right)}},
	\ee
if $\Delta t~m \le \pi\left[\tilde{x}^{2}(t)+y^{2}(t)\right]$, and $\lambda=0$ otherwise.  The time step does not necessarily need to be small; it can be chosen in the same way as in the parallel-shear and blinking vortex-vortex flows.

The parameter $m$ is the `pull' strength of the sink. If $m$ is very large, every node will be inevitably pulled to the sink during the half period in which the sink is active. The entire population will then be concentrated at $\left(x_s,0\right)$. For sufficiently small $m$, more interesting dynamics are obtained. In our simulations we use $m=0.03$ and $x_{s}=0.25$. 

As in the Blinking Vortex-Vortex flow, the domain in which the particles move is not bounded. However, we find that the stationary density is restricted to a finite area.
\label{LastPageApp}		


\end{document}